\documentclass{ptptex}
\usepackage{graphicx}

\newcommand{\eqdef}{\stackrel{\rm def}{=}}
\newcommand{\n}{\nonumber \\}
\newcommand{\bm}{\boldsymbol}

\title{
\vspace*{-20mm}
\begin{flushright}
  {\rm \small DPSU-05-2, September 2005}\\
  {\rm \small Prog.Theor.Phys.{\bf 114}(2005)1245-1260}\\
  {\rm \small hep-th/0512155}
\end{flushright}
\medskip
Calogero-Sutherland-Moser Systems,\\
Ruijsenaars-Schneider-van Diejen Systems and\\
Orthogonal Polynomials
}

\author{
Satoru \textsc{Odake}$^{1}$ and Ryu \textsc{Sasaki}$^{2}$
}

\inst{
$^1$Department of Physics, Shinshu University,
Matsumoto 390-8621, Japan\\
$^2$Yukawa Institute for Theoretical Physics, Kyoto University,\\
Kyoto 606-8502, Japan
}

\abst{
The equilibrium positions of the multi-particle classical
Calogero-Sutherland-Moser (CSM) systems with rational/trigonometric 
potentials associated with the classical root systems are described by 
the classical orthogonal polynomials; the Hermite, Laguerre and Jacobi
polynomials. The eigenfunctions of the corresponding single-particle
quantum CSM systems are also expressed in terms of
the same orthogonal polynomials.
We show that this interesting property is inherited by the 
Ruijsenaars-Schneider-van Diejen (RSvD) systems, which are integrable
deformation of the CSM systems; the equilibrium positions of the multi-particle
classical RSvD systems and the eigenfunctions of the corresponding 
single-particle quantum RSvD systems are described by the same orthogonal 
polynomials, the continuous Hahn (special case), Wilson and Askey-Wilson
polynomials. 
They belong to the Askey-scheme of the basic hypergeometric
orthogonal polynomials and are deformation of the Hermite, Laguerre and
Jacobi polynomials, respectively.
The Hamiltonians of these single-particle quantum mechanical systems
have two remarkable properties, factorization and shape invariance.
}

\begin{document}

\maketitle

\section{Introduction}

Exactly solvable and/or quasi-exactly solvable multi-particle
quantum mechanical systems have many remarkable properties.
Especially, those of the Calogero-Sutherland-Moser (CSM)
systems\cite{C71,S72,CM75} and their integrable deformation, called the
Ruijsenaars-Schneider-van Diejen (RSvD) systems,\cite{RS86,vD94} have
been extensively studied.
They have many attractive features at both classical and quantum
mechanical levels and are related to topics studied in other fields.

The equilibrium positions of the classical CSM systems with
rational/trigono\-met\-ric potentials associated with the classical root
systems 
are given by the zeros of the classical orthogonal polynomials, namely,
the Hermite, Laguerre and Jacobi (Chebyshev, Legendre, Gegenbauer) 
polynomials.\cite{C77,Sti,Sze,OS1} \  
For example, the Hamiltonian of the $A$-type multi-particle classical 
Calogero system is
\begin{equation}
  H=\sum_{j=1}^n\frac{1}{2m}p_j^2+
  \sum_{j=1}^n\frac12m\omega^2q_j^2
  +\frac{\bar{g}^2}{2m}\sum_{\genfrac{}{}{0pt}{2}{j,k=1}{j\neq k}}^n
  \frac{1}{(q_j-q_k)^2}+\text{const.}\,,
\end{equation}
and its equilibrium positions $\bar{q}_j$ 
($y_j=\sqrt{\frac{m\omega}{\bar{g}}}\bar{q}_j$) are determined by 
\begin{equation}
  \sum_{\genfrac{}{}{0pt}{2}{k=1}{k\neq j}}^n\frac{1}{y_j-y_k}=y_j\,.\qquad
  (j=1,\ldots,n) 
  \label{CalogeroEquiv}
\end{equation}
This equation can be converted into a differential equation for
$f(y)=\prod_{j=1}^n(y-y_j)$.
It is easy to show that $\frac{f^{\prime\,\prime}}{f}-2y\frac{f'}{f}
\xrightarrow{y\to y_j}\text{const}$. Thus the polynomial $f(y)$ satisfies
$f^{\prime\,\prime}-2yf'=Af$, in which the constant $A$ is determined as
$A=-2n$. This is the differential equation for the Hermite
polynomial $H_n(y)$. Therefore the equilibrium positions are given by
the zeros of the Hermite polynomial.
On the other hand, the corresponding single-particle
quantum system is described by the Hamiltonian of 
\begin{equation}
  H=\frac{1}{2m}p^2+\frac12m\omega^2q^2+\text{const.}
\end{equation}
This is just a harmonic oscillator, and its eigenfunctions are given by
(ground state eigenfunction) $\times$ (Hermite polynomials).
Interestingly, the equilibrium positions of the multi-particle classical
system and the eigenfunctions of the corresponding single-particle
quantum system are described by the same orthogonal polynomials, the Hermite
polynomials. This is not a fortuitous coincidence.
In fact, the $B$-type (or $C$-type) Calogero
systems and the $BC$-type Sutherland systems have the same property. In
these cases, the relevant orthogonal polynomials are the Laguerre and Jacobi
polynomials, respectively.

In this article, we establish that the above interesting property is
inherited by the RSvD systems, which are integrable deformation of the
CSM systems. Namely, we show that
the equilibrium positions of the multi-particle classical RSvD
systems with rational/trigonometric potentials associated with the
classical root systems and the eigenfunctions of the corresponding
single-particle quantum RSvD systems are described by the same
orthogonal polynomials; the continuous Hahn (special case),
Wilson and Askey-Wilson polynomials, which are deformation of the
Hermite, Laguerre and Jacobi polynomials, respectively. 
These polynomials are members of the Askey-scheme of the basic
hypergeometric orthogonal polynomials.\cite{AW85,KS96,AAR99} \ 
Deformation patterns are shown in Table I.
For the trigonometric $A_{n-1}$ systems, the situation is rather trivial;
the equilibrium positions are equally-spaced and
translationally invariant, 
and the single-particle system is a free theory. Although we are able to
relate them to Chebyshev polynomials of the first kind,
\cite{OS1,OS2,OSrokko} \ we do not demonstrate this here.
\begin{table}
\caption{Deformation patterns.}
\begin{center}
\begin{tabular}{cclcl}
  \hline \hline
  potential&root system&CSM system&$\rightsquigarrow$&RSvD system\\
  \hline
  rational&$A_{n-1}$&Hermite poly.&$\rightsquigarrow$&
  continuous Hahn poly.\\
  rational&$BC_n$&Laguerre poly.&$\rightsquigarrow$&Wilson poly.\\
  trigonometric&$BC_n$&Jacobi poly.&$\rightsquigarrow$&Askey-Wilson poly.\\
  \hline
\end{tabular}
\end{center}
\end{table}

For these single-particle quantum mechanical systems, the Hamiltonian
has two remarkable properties, {\em factorization} and
{\em shape invariance}.\cite{IH51,Crum55,Genden,SVZ93,susyqm} \ 
Shape invariance is an important ingredient of many exactly solvable
quantum mechanics.
In our case, the shape invariance determines the eigenfunctions and spectrum
from the data of the ground state eigenfunction and the energy of the
first excited state.\cite{OS4,OS5}

This article is organized as follows.
In \S 2 we recapitulate the essence of the CSM systems with
rational/trigonometric potentials associated with the classical root
systems. We briefly outline the argument 
that the equilibrium positions 
of the multi-particle classical systems and the eigenfunctions of the
corresponding single-particle quantum systems are described by the same
orthogonal polynomials; the Hermite, Laguerre and Jacobi polynomials.
In \S 3 we generalize this argument to the RSvD systems. The Hermite, Laguerre
and Jacobi polynomials are deformed to the continuous Hahn (special case),
Wilson and Askey-Wilson polynomials, respectively.
Section 4 is for a summary and comments.
In order to study single-particle systems, we repeatedly use the shape 
invariance of the single-particle Hamiltonian. 
In appendix some useful techniques for
shape invariant single-particle (`discrete') quantum
mechanics are summarized along the idea of Crum.\cite{Crum55}

\section{Calogero-Sutherland-Moser systems}

In this section we summarize the basics of the Calogero-Sutherland-Moser
systems with rational/trigonometric potentials associated with the classical
root systems.

\subsection{Models}

A CSM system is a multi-particle quantum (or classical) mechanical system
governed by a Hamiltonian $H(p,q)$ (or classical one, $H^{\rm class}(p,q)$).
\cite{C71,S72,CM75} \ 
The dynamical variables are the real-valued coordinates
$q={}^t(q_1,\cdots,q_n)$ and their canonically conjugate momenta
$p={}^t(p_1,\cdots,p_n)$.
For the quantum case we have $p_j=-i\hbar\frac{\partial}{\partial q_j}$.
We keep dimensionful parameters, e.g., mass, angular frequency, the
Planck constant, etc. The coordinate $q_j$ has the dimension of length.

The Hamiltonian of the CSM systems is
\begin{equation}
  H_{\rm CS}(p,q)=\sum_{j=1}^n\frac{1}{2m}p_j^2+V_{\rm CS}(q)\,,
  \label{H_CS}
\end{equation}
where the potential $V_{\rm CS}(q)$ can be written in terms of the
prepotential $W(q)$,
\begin{equation}
  V_{\rm CS}(q)=\sum_{j=1}^n\frac{1}{2m}\biggl(
  \Bigl(\frac{\partial W(q)}{\partial q_j}\Bigr)^2
  +\hbar\,\frac{\partial^2 W(q)}{\partial q_j^2}\biggr)\,.
  \label{V=W^2+dW}
\end{equation}
The explicit forms of the potential $V_{\rm CS}(q)$ and the prepotential
$W(q)$ are as follows:\\
(\romannumeral1) rational $A_{n-1}$ :
\begin{subequations}
\begin{align}
  V_{\rm CS}(q)&=\sum_{j=1}^n\frac12m\omega^2q_j^2
  +\frac{\hbar^2}{2m}\sum_{\genfrac{}{}{0pt}{2}{j,k=1}{j\neq k}}^n
  \frac{g(g-1)}{(q_j-q_k)^2}
  -\tfrac12\hbar\omega n\bigl(1+g(n-1)\bigr)\,,
  \label{ratA_CS_V}\\
  W(q)&=-\sum_{j=1}^n\frac12m\omega q_j^2+\sum_{1\leq j<k\leq n}
  g\hbar\log\sqrt{\tfrac{m\omega}{\hbar}}\,\bigl|q_j-q_k\bigr|\,,
  \label{ratA_CS_W}
\end{align}
\end{subequations}
(\romannumeral2) rational $BC_n$ :\,\footnote{
Since the independent coupling constants are $g_M$ and $g_S+g_L$,
this $BC_n$ model is equivalent to the $B_n$ model (or the $C_n$ model).
}
\begin{subequations}
\begin{align}
  V_{\rm CS}(q)&=\sum_{j=1}^n\biggl(\frac12m\omega^2q_j^2
  +\frac{\hbar^2}{2m}\frac{(g_S+g_L)(g_S+g_L-1)}{q_j^2}\biggr)\nonumber
  \displaybreak[3]\\
  &\quad
  +\frac{\hbar^2}{2m}\sum_{\genfrac{}{}{0pt}{2}{j,k=1}{j\neq k}}^n
  \biggl(\frac{g_M(g_M-1)}{(q_j-q_k)^2}+\frac{g_M(g_M-1)}{(q_j+q_k)^2}
  \biggr)\n
  &\quad
  -\hbar\omega n\bigl(g_S+g_L+\tfrac12+g_M(n-1)\bigr)\,,
  \label{ratBC_CS_V}\\
  W(q)&=-\sum_{j=1}^n\frac12m\omega q_j^2+\sum_{1\leq j<k\leq n}
  g_M\hbar\Bigl(\log\sqrt{\tfrac{m\omega}{\hbar}}\,\bigl|q_j-q_k\bigr|
  +\log\sqrt{\tfrac{m\omega}{\hbar}}\,\bigl|q_j+q_k\bigr|\Bigr)\n
  &\quad+\sum_{j=1}^n\Bigl(
  g_S\hbar\log\sqrt{\tfrac{m\omega}{\hbar}}\,\bigl|q_j\bigr|
  +g_L\hbar\log\sqrt{\tfrac{m\omega}{\hbar}}\,\bigl|2q_j\bigr|\Bigr)\,,
  \label{ratBC_CS_W}
\end{align}
\end{subequations}
(\romannumeral3) trigonometric $BC_n$ :
\begin{subequations}
\begin{align}
  V_{\rm CS}(q)&=\frac{\hbar^2\pi^2}{2mL^2}\sum_{j=1}^n\biggl(
  \frac{(g_S+g_L)(g_S+g_L-1)}{\sin^2\frac{\pi}{L}q_j}
  +\frac{g_L(g_L-1)}{\cos^2\frac{\pi}{L}q_j}\biggl)\n
  &\quad
  +\frac{\hbar^2\pi^2}{2mL^2}\sum_{\genfrac{}{}{0pt}{2}{j,k=1}{j\neq k}}^n
  \biggl(\frac{g_M(g_M-1)}{\sin^2\frac{\pi}{L}(q_j-q_k)}
  +\frac{g_M(g_M-1)}{\sin^2\frac{\pi}{L}(q_j+q_k)}\biggr)\n
  &\quad
  -\frac{\hbar^2\pi^2}{2mL^2}n\Bigl(\bigl(g_S+2g_L+g_M(n-1)\bigr)^2
  +g_M^2\tfrac13(n^2-1)\Bigr)\,,
  \label{trigBC_CS_V}\\
  W(q)&=\sum_{1\leq j<k\leq n}g_M\hbar\Bigl(
  \log\bigl|\sin\tfrac{\pi}{L}(q_j-q_k)\bigr|
  +\log\bigl|\sin\tfrac{\pi}{L}(q_j+q_k)\bigr|\Bigr)\n
  &\quad
  +\sum_{j=1}^n\Bigl(g_S\hbar\log\bigl|\sin\tfrac{\pi}{L}q_j\bigr|
  +g_L\hbar\log\bigl|\sin\tfrac{\pi}{L}2q_j\bigr|\Bigr)\,.
  \label{trigBC_CS_W}
\end{align}
\end{subequations}
The constant terms in $V_{\rm CS}(q)$ are the consequences of the
expression (\ref{V=W^2+dW}) in terms of the prepotential.
A constant shift of $W(q)$ does not affect \eqref{V=W^2+dW}.
In the above formulas $g,g_S,g_M$ and $g_L$ are dimensionless coupling
constants and we assume they are positive.
The other notation is conventional:
$m$ is the mass of the particles, $\omega$ is the angular frequency,
$\hbar$ is the Planck constant (divided by $2\pi$) and $L$ is the
circumference. All these parameters are positive.

Note that the Hamiltonian \eqref{H_CS} can be expressed as a sum of
factorized forms:
\begin{equation}
  H_{\rm CS}=\sum_{j=1}^n\frac{1}{2m}
  \Bigl(p_j-i\,\frac{\partial W(q)}{\partial q_j}\Bigr)
  \Bigl(p_j+i\,\frac{\partial W(q)}{\partial q_j}\Bigr)\,.
\end{equation}

\subsection{Equilibrium positions of $n$-particle classical mechanics}
\label{CSequiv}

The classical Hamiltonian $H^{\rm class}(p,q)$ is obtained from the quantum
one $H(p,q)$ by the following procedure: (a) regard $p_j$ is a $c$-number;
(b) after replacing the dimensionless coupling constants $g,g_1,g_2,\cdots$ by
the dimensionful ones $\bar{g}=g\hbar, {\bar{g}}_1=g_1\hbar,
{\bar{g}}_2=g_2\hbar,\cdots$, assume
$\bar{g},{\bar{g}}_1,{\bar{g}}_2,\cdots$ are independent of $\hbar$ ;
(c) take the $\hbar\rightarrow0$ limit.
In the same way, $V_{\rm CS}^{\rm class}(q)$ and $W^{\rm class}(q)$
are also obtained.

The canonical equations of motion of the classical systems are
\begin{equation}
  \frac{dq_j}{dt}=\frac{\partial H^{\rm class}(p,q)}{\partial p_j}\,,\quad
  \frac{dp_j}{dt}=-\frac{\partial H^{\rm class}(p,q)}{\partial q_j}\,.
\end{equation}
The equilibrium positions are given by the stationary solution
\begin{equation}
  p=0\,,\quad q=\bar{q}\,,
  \label{stasol}
\end{equation}
in which $\bar{q}$ satisfies
\begin{equation}
  \frac{\partial H^{\rm class}(0,q)}{\partial q_j}\Biggm|_{q=\bar{q}}=0\,.
  \qquad(j=1,\ldots,n)
  \label{staeq}
\end{equation}

For the CSM system, \eqref{staeq} becomes
$\frac{\partial V^{\rm class}_{\rm CS}(q)}{\partial q_j}\Bigm|_{q=\bar{q}}=0$
and it is equivalent to the condition \cite{CorS02}
\begin{equation}
  \frac{\partial W^{\rm class}(q)}{\partial q_j}\Biggm|_{q=\bar{q}}=0\,.
  \qquad(j=1,\ldots,n)
  \label{CS_equiveqW}
\end{equation}
For the rational $A_{n-1}$ system, the above condition reduces to
\eqref{CalogeroEquiv}.
Let us consider a polynomial whose zeros give the equilibrium positions:
\begin{subequations}
\begin{alignat}{2}
  &\text{(\romannumeral1) rational $A_{n-1}$}\,:&
  f_n(y)&=
  \prod_{j=1}^n\Bigl(y-\sqrt{\frac{m\omega}{\bar{g}}}\,\bar{q}_j\Bigr)
  \,,\\
  &\text{(\romannumeral2) rational $BC_n$}\,:&
  f_n(y)&=
  \prod_{j=1}^n\Bigl(y^2-\frac{m\omega}{\bar{g}_M}\,\bar{q}_j^2\Bigr)
  \,,\\
  &\text{(\romannumeral3) trigonometric $BC_n$}\,:\quad&
  f_n(\xi)&=
  \prod_{j=1}^n\Bigl(\xi-\cos\bigl(2\tfrac{\pi}{L}\bar{q}_j\bigr)\Bigr)
  \,.
\end{alignat}
\end{subequations}
Then Eq.\eqref{CS_equiveqW} can be converted into a differential
equation for $f_n(x)$, which is satisfied
by the Hermite, Laguerre and Jacobi polynomials, respectively.
The results are (see, for example, Refs.~\citen{C77,Sze,CorS02,OS2})
\begin{subequations}
\begin{alignat}{2}
  \text{(\romannumeral1)}\,&:&
  f_n(y)&=H_n^{\rm monic}(y)\,,\\
  \text{(\romannumeral2)}\,&:&
  f_n(y)&=L_n^{(\alpha)\,{\rm monic}}(y^2)\,,\quad
  \alpha=\frac{\bar{g}_S+\bar{g}_L}{\bar{g}_M}-1\,,\\
  \text{(\romannumeral3)}\,&:&\quad
  f_n(\xi)&=P_n^{(\alpha,\beta)\,{\rm monic}}(\xi)\,,\quad
  \alpha=\frac{\bar{g}_S+\bar{g}_L}{\bar{g}_M}-1\,,\ 
  \beta=\frac{\bar{g}_L}{\bar{g}_M}-1\,.
\end{alignat}
\end{subequations}
Here $H_n(y)\!=\!2^nH_n^{\rm monic}(y)$,
$L_n^{(\alpha)}(y^2)\!=\!\frac{(-1)^n}{n!}L_n^{(\alpha)\,{\rm monic}}(y^2)$
and $P_n^{(\alpha,\beta)}(\xi)\!=\!\frac{1}{2^n}\binom{\alpha+\beta+2n}{n}$
$P_n^{(\alpha,\beta)\,{\rm monic}}(\xi)$
are the Hermite, Laguerre and Jacobi polynomials, respectively.\cite{KS96}

\subsection{Eigenfunctions of single-particle quantum mechanics}

The Hamiltonian \eqref{H_CS} for single-particle case ($n=1$) reads
\begin{subequations}
\begin{alignat}{2}
  \text{(\romannumeral1)}\,&:&
  H&=-\frac{\hbar^2}{2m}\frac{d^2}{dx^2}+\frac12m\omega^2x^2
  -\frac12\hbar\omega\,,\\
  \text{(\romannumeral2)}\,&:&
  H&=-\frac{\hbar^2}{2m}\frac{d^2}{dx^2}+\frac12m\omega^2x^2
  +\frac{\hbar^2}{2m}\frac{g(g-1)}{x^2}-\hbar\omega\Bigl(g+\frac12\Bigr)\,,\\
  \text{(\romannumeral3)}\,&:&\quad
  H&=-\frac{\hbar^2}{2m}\frac{d^2}{dx^2}
  +\frac{\hbar^2\pi^2}{2mL^2}\Bigl(
  \frac{g(g-1)}{\sin^2\frac{\pi}{L}x}+\frac{g'(g'-1)}{\cos^2\frac{\pi}{L}x}
  \Bigr)
  -\frac{\hbar^2\pi^2}{2mL^2}(g+g')^2\,,
\end{alignat}
\end{subequations}
where $x=q_1$. These are
a harmonic oscillator, a harmonic oscillator
with a centrifugal barrier, and the P\"{o}schl-Teller potential,\cite{PT33} \ 
respectively, with some constant energy shifts.
By introducing the dimensionless variable $y$,
\begin{equation}
  \text{(\romannumeral1), (\romannumeral2)}\,:\quad
  y=\sqrt{\frac{m\omega}{\hbar}}x\,,\qquad
  \text{(\romannumeral3)}\,:\quad
  y=\frac{\pi}{L}x\,, 
\end{equation}
these Hamiltonians are reduced to dimensionless ones $\mathcal{H}$:
\begin{subequations}
\begin{alignat}{3}
  \text{(\romannumeral1)}\,&:&
  H&=\hbar\omega\mathcal{H}\,,&
  \mathcal{H}&=-\frac12\frac{d^2}{dy^2}+\frac12y^2-\frac12\,,
  \label{ratA_CS_cH}\\
  \text{(\romannumeral2)}\,&:&
  H&=\hbar\omega\mathcal{H}\,,&
  \mathcal{H}&=-\frac12\frac{d^2}{dy^2}+\frac12y^2
  +\frac{g(g-1)}{2y^2}-g-\frac12\,,\\
  \text{(\romannumeral3)}\,&:&\quad
  H&=\frac{\hbar^2\pi^2}{mL^2}\,\mathcal{H}\,,&\quad
  \mathcal{H}&=\frac12\Bigl(-\frac{d^2}{dy^2}+\frac{g(g-1)}{\sin^2y}
  +\frac{g'(g'-1)}{\cos^2y}-(g+g')^2\Bigr)\,.
\end{alignat}
\end{subequations}
They have two properties, {\em factorization} \eqref{rat_CS_calH=AA}
and {\em shape invariance} \eqref{rat_CS_shapeinv}.
In the notation of the appendix, the necessary data of the shape invariance
are the invariant (pre)potential $\mathcal{W}(y\,;\bm{\lambda})$, the
set of parameters $\bm{\lambda}$, the elementary shift of the parameters
$\bm{\delta}$ and the energy of the first excited state 
$\mathcal{E}_1(\bm{\lambda})$:
\begin{subequations}
\begin{align}
  \text{(\romannumeral1)}\,:\,\quad&
  \mathcal{W}(y)=-\frac12y^2,\quad 
  \text{no $\bm{\lambda}$ (no $\bm{\delta}$)},\quad
  \mathcal{E}_1=1,\\
  \text{(\romannumeral2)}\,:\,\quad&
  \mathcal{W}(y\,;\bm{\lambda})=-\tfrac12y^2+g\log y\,,\quad
  \bm{\lambda}=g,\quad \bm{\delta}=1,\quad
  \mathcal{E}_1(\bm{\lambda})=2,\\
  \text{(\romannumeral3)}\,:\,\quad&
  \mathcal{W}(y\,;\bm{\lambda})=g\log\sin y+g'\log\cos y\,,\n
  &
  \bm{\lambda}=(g,g'),\quad \bm{\delta}=(1,1),\quad
  \mathcal{E}_1(\bm{\lambda})=2(g+g'+1).
\end{align}
\end{subequations}
{}From the formulas \eqref{rat_CS_phi=Adphi}--\eqref{rat_CS_phi=phi}
and \eqref{rat_CS_E_rec}, we obtain the eigenfunctions $P_n\phi_0$
and the corresponding eigenvalues $\mathcal{E}_n$ of $\mathcal{H}$:
\begin{subequations}
\begin{align}
  \text{(\romannumeral1)}\,:\,\quad&
  P_n(y)\propto H_n(y),\quad \mathcal{E}_n=n,\\
  \text{(\romannumeral2)}\,:\,\quad&
  P_n(y\,;\bm{\lambda})\propto L_n^{(g-\frac12)}(y^2),\quad
  \mathcal{E}_n(\bm{\lambda})=2n,\\
  \text{(\romannumeral3)}\,:\,\quad&
  P_n(y\,;\bm{\lambda})\propto P_n^{(g-\frac12,g'-\frac12)}(\cos 2y),\quad
  \mathcal{E}_n(\bm{\lambda})=2n(n+g+g').
\end{align}
\end{subequations}

Thus
the equilibrium positions of the multi-particle classical systems
and the eigenfunctions of the single-particle quantum systems are
described by the same orthogonal polynomials; the Hermite, Laguerre and
Jacobi polynomials.

\section{Ruijsenaars-Schneider-van Diejen systems}

In this section we generalize the results of the previous section to the
Ruijsenaars-Schneider-van Diejen (RSvD) systems with the rational/trigonometric
potentials associated with the classical root systems.

\subsection{Models}

The RSvD system is an integrable deformation of the CSM
system.\cite{RS86,vD94} \ 
The Hamiltonian of the RSvD systems is
\begin{equation}
  H(p,q)=\frac12mc^2\sum_{j=1}^n\Bigl(
  \sqrt{V_j(q)}\,e^{\frac{1}{mc}p_j}\sqrt{{V_j(q)}^*}
  +\sqrt{{V_j(q)}^*}\,e^{-\frac{1}{mc}p_j}\sqrt{V_j(q)}
  -V_j(q)-{V_j(q)}^*\Bigr)\,,
  \label{H_RS}
\end{equation}
where $V_j(q)$ is given by
\begin{equation}
  V_j(q)=w(q_j)\prod_{\genfrac{}{}{0pt}{2}{k=1}{k\neq j}}^n
  v(q_j-q_k)\times
  \begin{cases}
  1&\text{for $A_{n-1}$,}\\v(q_j+q_k)&\text{for $BC_n$}\,.
  \end{cases}
\end{equation}
We use the conventional notation that $V_j(q)^*$ is the complex
conjugate of $V_j(q)$.
Since the operators $e^{\pm\frac{1}{mc}p_j}=
e^{\mp i\frac{\hbar}{mc}\frac{\partial}{\partial q_j}}$ cause finite
shifts of the wavefunction in the imaginary direction
($e^{\pm\frac{1}{mc}p_j}f(q)=
f(q_1,\cdots,q_j\mp i\frac{\hbar}{mc},\cdots,q_n)$),
we call these systems `discrete' dynamical systems. (Sometimes they are 
misleadingly called `relativistic' version of the CSM.
See Ref.~\citen{BS97} for comments on this point.)
The basic potential functions $v(x)$ and $w(x)$ are as follows:\\
(\romannumeral1) rational $A_{n-1}$ :
\begin{subequations}
\begin{align}
  v(x)&=1-i\frac{\hbar}{mc}\frac{g}{x}\,,
  \label{ratA_RS_v}\\
  w(x)&=\Bigl(1+i\,\frac{\omega_1}{c}\,x\Bigr)
  \Bigl(1+i\,\frac{\omega_2}{c}\,x\Bigr)\,,
  \label{ratA_RS_w}
\end{align}
\end{subequations}
(\romannumeral2) rational $BC_n$ :
\begin{subequations}
\begin{align}
  v(x)&=1-i\frac{\hbar}{mc}\frac{g_0}{x}\,,
  \label{ratBC_RS_v}\\
  w(x)&=\Bigl(1+i\,\frac{\omega_1}{c}\,x\Bigr)
  \Bigl(1+i\,\frac{\omega_2}{c}\,x\Bigr)
  \Bigl(1-i\frac{\hbar}{mc}\frac{g_1}{x}\Bigr)
  \Bigl(1-i\frac{\hbar}{mc}\frac{g_2}{x-i\frac{\hbar}{2mc}}\Bigr)\,,
  \label{ratBC_RS_w}
\end{align}
\end{subequations}
(\romannumeral3) trigonometric $BC_n$ :
\begin{subequations}
\begin{align}
  v(x)&=\frac{\sin\frac{\pi}{L}(x-i\frac{\hbar}{mc}g_0)}{\sin\frac{\pi}{L}x}
  \,,
  \label{trigBC_RS_v}\\
  w(x)&=
  \frac{\sin\frac{\pi}{L}(x-i\frac{\hbar}{mc}g_1)}{\sin\frac{\pi}{L}x}\,
  \frac{\sin\frac{\pi}{L}(x-i\frac{\hbar}{2mc}-i\frac{\hbar}{mc}g_2)}
       {\sin\frac{\pi}{L}(x-i\frac{\hbar}{2mc})}\n
  &\quad\times
  \frac{\cos\frac{\pi}{L}(x-i\frac{\hbar}{mc}g'_1)}{\cos\frac{\pi}{L}x}\,
  \frac{\cos\frac{\pi}{L}(x-i\frac{\hbar}{2mc}-i\frac{\hbar}{mc}g'_2)}
       {\cos\frac{\pi}{L}(x-i\frac{\hbar}{2mc})}\,.
  \label{trigBC_RS_w}
\end{align}
\end{subequations}
Here $g,g_0,g_1,g_2,g'_1$ and $g'_2$ are dimensionless coupling
constants and $c$ is the (fictitious) speed of light. 
We assume they are all positive.

Note that the Hamiltonian \eqref{H_RS} can be expressed as a sum of
factorized forms,
\begin{equation}
  H=\frac12mc^2\sum_{j=1}^n
  \Bigl(\sqrt{V_j(q)}\,e^{\frac{1}{2mc}p_j}
        -\sqrt{{V_j(q)}^*}\,e^{-\frac{1}{2mc}p_j}\Bigr)
  \Bigl(e^{\frac{1}{2mc}p_j}\sqrt{{V_j(q)}^*}
        -e^{-\frac{1}{2mc}p_j}\sqrt{V_j(q)}\,\Bigr)\,.
  \label{H_RS_factor}
\end{equation}
We also remark that in the $c\rightarrow\infty$ limit, the RSvD systems
reduce to the CSM systems,
\begin{equation}
  \lim_{c\rightarrow\infty}H(p,q)=H_{\rm CS}(p,q),
  \label{RSCS}
\end{equation}
where the correspondence of the parameters is
\begin{subequations}
\begin{align}
  \text{(\romannumeral1)}\,:\,\quad&
  \omega_1+\omega_2=\omega,\quad g=g\,,
  \label{ratA_RSCS_para}\\
  \text{(\romannumeral2)}\,:\,\quad&
  \omega_1+\omega_2=\omega,\quad g_0=g_M,\quad g_1+g_2=g_S+g_L\,,
  \label{ratBC_RSCS_para}\\
  \text{(\romannumeral3)}\,:\,\quad&
  g_0=g_M,\quad g_1+g_2=g_S+g_L,\quad g'_1+g'_2=g_L\,.
  \label{trigBC_RSCS_para}
\end{align}
\end{subequations}

\subsection{Equilibrium positions of $n$-particle classical mechanics}

The classical Hamiltonian $H^{\rm class}(p,q)$ (and also $V_j^{\rm class}(q)$, 
$v^{\rm class}(x)$, $w^{\rm class}(x)$) is obtained by the
prescription given in \S\ref{CSequiv}.
For the RSvD system, the equation for the equilibrium positions \eqref{staeq} 
is equivalent to the condition \cite{RagS04}
\begin{equation}
  V_j^{\rm class}(\bar{q})=V_j^{\rm class}(\bar{q})^*>0\,.\qquad(j=1,\ldots,n)
  \label{RS_equiveq}
\end{equation}
This equation {\em without the inequality sign\/} can be rewritten
in a Bethe-ansatz-like form,
\begin{equation}
  \prod_{\genfrac{}{}{0pt}{2}{k=1}{k\neq j}}^n
  \frac{v^{\rm class}(\bar{q}_j-\bar{q}_k)\,v^{\rm class}(\bar{q}_j+\bar{q}_k)}
  {v^{\rm class}(\bar{q}_j-\bar{q}_k)^*\,v^{\rm class}(\bar{q}_j+\bar{q}_k)^*}
  =\frac{w^{\rm class}(\bar{q}_j)^*}{w^{\rm class}(\bar{q}_j)}\,.
  \qquad(j=1,\ldots,n)
  \label{equiveqBA}
\end{equation}
(For the $A_{n-1}$ type systems, $v^{\rm class}(\bar{q}_j+\bar{q}_k)$ and
$v^{\rm class}(\bar{q}_j+\bar{q}_k)^*$ should be omitted.)

Let us consider a polynomial whose zeros give the equilibrium positions:
\begin{subequations}
\begin{alignat}{2}
  &\text{(\romannumeral1) rational $A_{n-1}$}\,:&
  f_n(y)&=
  \prod_{j=1}^n\Bigl(y-\sqrt{\frac{m\omega_1}{\bar{g}}}\,\bar{q}_j\Bigr)
  \,,\\
  &\text{(\romannumeral2) rational $BC_n$}\,:&
  f_n(y)&=
  \prod_{j=1}^n\Bigl(y^2-\frac{m\omega_1}{\bar{g}_0}\,\bar{q}_j^2\Bigr)
  \,,\\
  &\text{(\romannumeral3) trigonometric $BC_n$}\,:\quad&
  f_n(\xi)&=
  \prod_{j=1}^n\Bigl(\xi-\cos\bigl(2\tfrac{\pi}{L}\bar{q}_j\bigr)\Bigr)
  \,.
\end{alignat}
\end{subequations}
Then Eq.\eqref{RS_equiveq} can be converted into a functional
equation for $f_n(x)$.\cite{OS2,OS5} \  For the simplest case
(\romannumeral1) with $\omega_2=0$, this functional equation can be
solved explicitly by using its generating function.
However, such a straightforward solution of the functional equation
cannot be easily obtained in the other cases.
For this reason, we solved \eqref{RS_equiveq}
numerically and guessed a recursion relation for $\{f_n(x)\}$.
We hypothesize that the polynomials 
$\{f_n(x)\}$ satisfy the three-term recurrence
$f_{n+1}(x)-(x-a_n)f_n(x)+b_nf_{n-1}(x)=0$ 
($n\geq 0$, $f_{-1}(x)=0$, $f_0(x)=1$) for some constants $a_n$ and $b_n$.
The three-term recurrence implies that $\{f_n(x)\}$ are orthogonal polynomials.
Then we can show that the polynomials $\{f_n(x)\}$ with this three-term
recurrence solve the functional equation. The results are 
\cite{OS2,OS5} (see also Refs.~\citen{RagS04,vD04,ILR04})
\begin{subequations}
\begin{align}
  \text{(\romannumeral1)}\,:\,\quad&
  f_n(y)
  =p_n^{\rm monic}\Bigl(\sqrt{\frac{mc^2}{\omega_1\bar{g}}}\,y\,;
  \frac{mc^2}{\omega_1\bar{g}},\frac{mc^2}{\omega_2\bar{g}},
  \frac{mc^2}{\omega_1\bar{g}},\frac{mc^2}{\omega_2\bar{g}}\Bigr)\,,\\
  \text{(\romannumeral2)}\,:\,\quad&
  f_n(y)
  =W_n^{\rm monic}\Bigl(\frac{mc^2}{\omega_1\bar{g}_0}\,y^2;
  \frac{mc^2}{\omega_1\bar{g}_0},\frac{mc^2}{\omega_2\bar{g}_0},
  \frac{\bar{g}_1}{\bar{g}_0},\frac{\bar{g}_2}{\bar{g}_0}\Bigr)\,,\\
  \text{(\romannumeral3)}\,:\,\quad&
  f_n(\xi)
  =p_n^{\rm monic}\bigl(\xi\,;
  e^{-\frac{2\pi\bar{g}_1}{mcL}},e^{-\frac{2\pi\bar{g}_2}{mcL}},
  -e^{-\frac{2\pi\bar{g}'_1}{mcL}},-e^{-\frac{2\pi\bar{g}'_2}{mcL}}
  \bigm|e^{-\frac{2\pi\bar{g}_0}{mcL}}\bigr)\,.
\end{align}
\end{subequations}
With proper normalization, these orthogonal polynomials are known 
as:\cite{KS96}
\begin{subequations}
\begin{align}
  \text{(\romannumeral1)}\,:\,\quad&
  \text{the continuous Hahn polynomials},\\
  &p_n(y;a_1,a_2,b_1,b_2)=\frac{1}{n!}(n+a_1+a_2+b_1+b_2-1)_n\,
  p_n^{\rm monic}(y;a_1,a_2,b_1,b_2),\n
  \text{(\romannumeral2)}\,:\,\quad&
  \text{the Wilson polynomials},\\
%
  &W_n(y^2;a_1,a_2,a_3,a_4)=
  (-1)^n(n+{\textstyle \sum_{\alpha=1}^4}a_{\alpha}-1)_n\,
  W_n^{\rm monic}(y^2;a_1,a_2,a_3,a_4),\n
  \text{(\romannumeral3)}\,:\,\quad&\text{the Askey-Wilson polynomials,}\\
  &p_n(\xi\,;a_1,a_2,a_3,a_4|q)=2^n(a_1a_2a_3a_4q^{n-1};q)_n\,
  p_n^{\rm monic}(\xi\,;a_1,a_2,a_3,a_4|q).\nonumber
\end{align}
\end{subequations}
These orthogonal polynomials are deformation of the Hermite, Laguerre
and Jacobi polynomials, respectively.

\subsection{Eigenfunctions of single-particle quantum mechanics}

The Hamiltonian \eqref{H_RS} for the single-particle case ($n=1$) reads
\begin{equation}
  H=\frac{mc^2}{2}\Bigl(
  \sqrt{w(x)}\,e^{-i\frac{\hbar}{mc}\frac{d}{dx}}\sqrt{w(x)^*}
  +\sqrt{w(x)^*}\,e^{i\frac{\hbar}{mc}\frac{d}{dx}}\sqrt{w(x)}
  -w(x)-w(x)^*\Bigr)\,,
  \label{rat_RS_H}
\end{equation}
where $x=q_1$. 
By introducing a dimensionless variable $y$ (or $z$), a rescaled
potential $V(y)$ (or $V(z)$), and a set of parameters $\bm{\lambda}$
(and a parameter $q$) given by
\begin{subequations}
\begin{align}
  \text{(\romannumeral1)}\,:\,\quad&
  y=\frac{mc}{\hbar}x\,,\quad\bm{\lambda}=(a_1,a_2),\quad
  V(y)=V\bigl(y\,;\bm{\lambda}\bigr)=(a_1+iy)(a_2+iy)\,,
  \label{y_lambda_V(i)}\\
  \text{(\romannumeral2)}\,:\,\quad&
  y=\frac{mc}{\hbar}x\,,\quad\bm{\lambda}=(a_1,a_2,a_3,a_4),\n
  &V(y)=V\bigl(y\,;\bm{\lambda}\bigr)
  =\frac{(a_1+iy)(a_2+iy)(a_3+iy)(a_4+iy)}{2iy(2iy+1)}\,,
  \label{ratBC_RS_V}\\
  \text{(\romannumeral3)}\,:\,\quad&
  y=\frac{\pi}{L}x,\quad z=e^{2iy}=e^{2\pi i\frac{x}{L}},\quad
  \bm{\lambda}=(a_1,a_2,a_3,a_4),\quad q=e^{-\frac{2\pi\hbar}{mcL}},\n
  &V(z)=V\bigl(z\,;\bm{\lambda},q\bigr)
  =\frac{(1-a_1z)(1-a_2z)(1-a_3z)(1-a_4z)}{(1-z^2)(1-qz^2)}\,,
  \label{trigBC_RS_V}
\end{align}
\end{subequations}
the function $w(x)$ takes the form
\begin{subequations}
\begin{align}
  \text{(\romannumeral1)}\,:\,\ &
  w(x)=\frac{1}{a_1a_2}V(y\,;\bm{\lambda})\,,\quad
  \bm{\lambda}=\Bigl(\frac{mc^2}{\hbar\omega_1},\frac{mc^2}{\hbar\omega_2}
  \Bigr)\,,\\
  \text{(\romannumeral2)}\,:\,\ &
  w(x)=\frac{4}{a_1a_2}V(y\,;\bm{\lambda})\,,\quad
  \bm{\lambda}=\Bigl(\frac{mc^2}{\hbar\omega_1},\frac{mc^2}{\hbar\omega_2},
  g_1,g_2+\tfrac12\Bigr)\,,\\
  \text{(\romannumeral3)}\,:\,\ &
  w(x)^*=(a_1a_2a_3a_4q^{-1})^{-\frac12}V(z\,;\bm{\lambda},q),\ \ 
  \bm{\lambda}=(q^{g_1},q^{g_2+\frac12},-q^{g'_1},-q^{g'_2+\frac12})\,.
  \label{w_lambda(iii)}
\end{align}
\end{subequations}
Then the Hamiltonian \eqref{rat_RS_H} reduces to the dimensionless one 
$\mathcal{H}$, with
\begin{subequations}
\begin{align}
  \text{(\romannumeral1)}\,:\,\quad&
  H=\frac{mc^2}{a_1a_2}\mathcal{H}\,,\quad
  \text{where $\mathcal{H}$ is given by \eqref{rat_RS_calH}}\,,\\
  \text{(\romannumeral2)}\,:\,\quad&
  H=\frac{4mc^2}{a_1a_2}\mathcal{H}\,,\quad
  \text{where $\mathcal{H}$ is given by \eqref{rat_RS_calH}}\,,\\
  \text{(\romannumeral3)}\,:\,\quad&
  H=mc^2(a_1a_2a_3a_4q^{-1})^{-\frac12}\mathcal{H}\,,\quad
  \text{where $\mathcal{H}$ is given by \eqref{trig_RS_calH}}\,.
\end{align}
\end{subequations}
These Hamiltonians have two properties, {\em factorization} and
{\em shape invariance}.\cite{OS4,OS5} \ 
In the notation used in the appendix, the necessary data for
the shape invariance
are the ground state wavefunction $\phi_0$, the elementary shift 
$\bm{\delta}$ of the parameters and the first energy level 
$\mathcal{E}_1(\bm{\lambda})$:
\begin{subequations}
\begin{align}
  \text{(\romannumeral1)}\,:\,\quad&
  \phi_0(y\,;\bm{\lambda})\propto
  \bigl|\Gamma(a_1+iy)\Gamma(a_2+iy)\bigr|\,,\quad
  \bm{\delta}=(\tfrac12,\tfrac12)\,,\quad
  \mathcal{E}_1(\bm{\lambda})=a_1+a_2\,,\\
  \text{(\romannumeral2)}\,:\,\quad&
  \phi_0(y\,;\bm{\lambda})\propto \biggl|
  \frac{\Gamma(a_1+iy)\Gamma(a_2+iy)\Gamma(a_3+iy)\Gamma(a_4+iy)}
  {\Gamma(2iy)}\biggr|\,,\n
  &\bm{\delta}=(\tfrac12,\tfrac12,\tfrac12,\tfrac12),\quad
  \mathcal{E}_1(\bm{\lambda})=\tfrac12(a_1+a_2+a_3+a_4)\,,\\
  \text{(\romannumeral3)}\,:\,\quad&
  \phi_0(z\,;\bm{\lambda},q)\propto \biggl|
  \frac{(z^2;q)_{\infty}}{(a_1z,a_2z,a_3z,a_4z;q)_{\infty}}\biggr|\,,\n
  &\delta=\frac12,\quad\delta'=-\frac12,\quad
  \mathcal{E}_1(\bm{\lambda},q)=\tfrac12(q^{-1}-1)(1-a_1a_2a_3a_4)\,.
\end{align}
\end{subequations}
{}From the formulas \eqref{rat_CS_phi=Adphi}--\eqref{rat_CS_phi=phi}/\eqref
{trig_RS_phi=Adphi}--\eqref{trig_RS_phi=phi}
and \eqref{rat_CS_E_rec}/\eqref{trig_RS_E_rec}, 
we obtain the eigenfunctions $P_n\phi_0$ and the corresponding
eigenvalues $\mathcal{E}_n$ of $\mathcal{H}$:
\begin{subequations}
\begin{align}
  \text{(\romannumeral1)}\,:\,\quad&
  P_n(y;\bm{\lambda})\propto p_n(y\,;a_1,a_2,a_1,a_2),\quad 
  \mathcal{E}_n(\bm{\lambda})=\tfrac12n(n+2a_1+2a_2-1)\,,\\
  \text{(\romannumeral2)}\,:\,\quad&
  P_n(y\,;\bm{\lambda})\propto W_n(y^2;a_1,a_2,a_3,a_4),\n
  &\mathcal{E}_n(\bm{\lambda})=\tfrac12n(n+a_1+a_2+a_3+a_4-1)\,,\\
  \text{(\romannumeral3)}\,:\,\quad&
  P_n(z;\bm{\lambda},q)\propto p_n({\rm Re}z\,;a_1,a_2,a_3,a_4|q),\n
  &\mathcal{E}_n(\bm{\lambda},q)=\tfrac12(q^{-n}-1)(1-a_1a_2a_3a_4q^{n-1})\,,
\end{align}
\end{subequations}
where the first $p_n$, $W_n$ and the second $p_n$ are the continuous Hahn, 
Wilson and Askey-Wilson polynomials in the notation used in Ref.~\citen{KS96}.

We thus find that 
the equilibrium positions of the multi-particle classical systems
and the eigenfunctions of the single-particle quantum systems are
described by the same orthogonal polynomials, namely, the continuous Hahn
(special case), Wilson and Askey-Wilson polynomials.

\section{Summary and comments}

We have established an interesting property of the
Ruijsenaars-Schneider-van Diejen systems with the
rational/trigonometric potentials associated with the classical root systems.
The equilibrium positions of the multi-particle classical systems and the 
eigenfunctions of the single-particle quantum systems are described by
the same orthogonal polynomials; the continuous Hahn (special case),
Wilson and Askey-Wilson polynomials.
This property is inherited from the Calogero-Sutherland-Moser systems,
in which the relevant orthogonal polynomials are the Hermite, Laguerre
and Jacobi polynomials.
These polynomials are members of the Askey-scheme of the basic
hypergeometric orthogonal polynomials.
The CSM and RSvD systems admit elliptic potentials and finding
eigenfunctions of such elliptic systems is a good challenge.
If this property is inherited by the elliptic RSvD systems, study of classical
equilibrium positions may shed light on the quantum problem of finding 
eigenfunctions, which is quite non-trivial.

This interesting property is obtained as a result of explicit computation
and at present we do not know any deeper reason or meaning behind it.
In \S 6 of Ref.~\citen{OS1}, we presented the `phenomenological' observation
that the Hermite, Laguerre and Jacobi polynomials of degree $n$ can be obtained
from the prepotential of CSM systems with $n+1$ particles.
Is this also true for the continuous Hahn, Wilson and
Askey-Wilson polynomials? Although there is no prepotential for the RSvD
systems, the corresponding quantity may be $\log\phi_0$.
The ground state eigenfunction $\phi_0(\{q_j\})$ of \eqref{H_RS} (see 
\eqref{H_RS_factor}) is
\begin{subequations}
\begin{align}
  \text{(\romannumeral1)}\,:\,\ &
  \phi_0(\{q_j\}\,;\bm{\lambda},g)\propto \biggl|\,
  \prod_{j=1}^n\Gamma(a_1+iy_j)\Gamma(a_2+iy_j)\,\cdot\!\!\!\!
  \prod_{1\leq j<k\leq n}\frac{\Gamma(g+i(y_j-y_k))}{\Gamma(i(y_j-y_k))}\,
  \biggr|\,,\\
  \text{(\romannumeral2)}\,:\,\ &
  \phi_0(\{q_j\}\,;\bm{\lambda},g_0)\propto \biggl|\, \prod_{j=1}^n  
  \frac{\prod_{\alpha=1}^4\Gamma(a_{\alpha}+iy_j)}
  {\Gamma(2iy_j)}\,\cdot\!\!\!\!
  \prod_{1\leq j<k\leq n}\prod_{\epsilon=\pm 1}
  \frac{\Gamma(g_0+i(y_j+\epsilon\,y_k))}{\Gamma(i(y_j+\epsilon\,y_k))}\,
  \biggr|\,,\\
  \text{(\romannumeral3)}\,:\,\ &
  \phi_0(\{q_j\}\,;\bm{\lambda},g_0,q)\propto \biggl|\, \prod_{j=1}^n
  \frac{(z_j^2;q)_{\infty}}{\prod_{\alpha=1}^4(a_{\alpha}z_j;q)_{\infty}}
  \,\cdot\!\!\!\!
  \prod_{1\leq j<k\leq n}\prod_{\epsilon=\pm 1}
  \frac{(z_jz_k^{\epsilon};q)_{\infty}}{(a_0z_jz_k^{\epsilon};q)_{\infty}}\,
  \biggr|\,,
\end{align}
\end{subequations}
where $\bm{\lambda}$ is given in \eqref{y_lambda_V(i)}--\eqref{w_lambda(iii)}
and $y_j=\frac{mc}{\hbar}q_j$, $z_j=e^{2\pi i\frac{q_j}{L}}$ and
$a_0=q^{g_0}$.
For the CSM systems, the second factor of $\phi_0$ is simply $q_j\pm q_j$
or $\sin(q_j\pm q_j)$. The observation of Ref.~\citen{OS1} is that the first
factor of $\phi_0$ gives the measure and the second factor gives the
desired polynomial. For the RSvD systems, the first factor of $\phi_0$ gives
the measure, but the second factor does not give the desired polynomial. 
Therefore, the observation of Ref.~\citen{OS1} does not apply to the
RSvD systems in its naivest form.

In the treatment of the Hamiltonians of these single-particle quantum
mechanics, we have emphasized factorization, shape invariance and
construction of the isospectral Hamiltonians. Although the examples
given in this article possess rational and trigonometric potentials, this
method and the basic idea can be applied to a wider class of potentials, e.g. 
elliptic potentials.
In ordinary quantum mechanics there exists Crum's theorem,\cite{Crum55} \ 
which describes the construction of the associated isospectral Hamiltonians
$\mathcal{H}_s$ and their eigenfunctions $\phi_{s,n}$ for general systems
without shape invariance. The construction of $\mathcal{H}_s$ and
$\phi_{s,n}$ given in this article for the `discrete' cases is based on shape
invariance. 
A `discrete' analogue of Crum's theorem, that is, the construction of 
the associated isospectral Hamiltonians and their eigenfunctions without
shape invariance, would be very helpful, if it exists.

\section*{Acknowledgements}

S.~O. and R.~S. are supported in part by Grant-in-Aid for Scientific
Research from the Ministry of Education, Culture, Sports, Science and
Technology, No. 13135205 and 14540259, respectively.

\appendix
\section*{Shape Invariance in Single-Particle (`Discrete') Quantum Mechanics}

In this appendix we collect useful techniques for the shape invariant
single-particle (`discrete') quantum mechanics along the idea of
Crum.\cite{Crum55} \ 
All the quantities in this appendix are dimensionless.

\subsection{Ordinary quantum mechanics}
\label{App:QM}

Let us consider an ordinary quantum mechanical system with the single 
coordinate $y$ governed by the Hamiltonian
\begin{equation}
  \mathcal{H}=-\frac12\frac{d^2}{dy^2}+\mathcal{V}(y),
\end{equation}
where the potential $\mathcal{V}$ may contain a set of parameters 
$\bm{\lambda}$. We assume that $\mathcal{H}$ has a square integrable ground
state and a discrete spectrum,
\begin{equation}
  \mathcal{H}\phi_n=\mathcal{E}_n\phi_n\quad(n=0,1,2,\ldots),\quad
  0=\mathcal{E}_0<\mathcal{E}_1<\mathcal{E}_2<\cdots,
  \label{cHphi=cEphi}
\end{equation}
where the constant term of $\mathcal{V}$ is chosen such that 
$\mathcal{E}_0=0$. Since the ground state eigenfunction $\phi_0$ has no
node, it can be expressed as $\phi_0(y)=e^{\mathcal{W}(y)}$,
and the excited state eigenfunctions are expressed as 
$\phi_n(y)=P_n(y)\phi_0(y)$.
Then the Hamiltonian $\mathcal{H}$ can be factorized as
\begin{align}
  \mathcal{H}&=\mathcal{A}(y)^{\dagger}\mathcal{A}(y)
  =\frac12\Bigl(-\frac{d^2}{dy^2}
  +\Bigl(\frac{d\mathcal{W}(y)}{dy}\Bigr)^2
  +\frac{d^2\mathcal{W}(y)}{dy^2}\Bigr)\,,
  \label{rat_CS_calH=AA}\\
  \mathcal{A}(y)&\eqdef\frac{1}{\sqrt{2}}\Bigl(
  -i\frac{d}{dy}+i\,\frac{d\,\mathcal{W}(y)}{dy}\Bigr)\,,\quad
  \mathcal{A}(y)^{\dagger}\eqdef\frac{1}{\sqrt{2}}\Bigl(
  -i\frac{d}{dy}-i\,\frac{d\,\mathcal{W}(y)}{dy}\Bigr)\,.
  \label{rat_CS_Ad}
\end{align}
The ground state $\phi_0$ is annihilated by $\mathcal{A}$,
$\mathcal{A}\phi_0=0$.

Let us write the parameter dependence explicitly, as
$\mathcal{A}(y)=\mathcal{A}(y\,;\bm{\lambda})$.
We assume the following {\em shape invariance}
\begin{equation}
  \mathcal{A}(y\,;\bm{\lambda})\mathcal{A}(y\,;\bm{\lambda})^{\dagger}
  =\mathcal{A}(y\,;\bm{\lambda}+\bm{\delta})^{\dagger}
  \mathcal{A}(y\,;\bm{\lambda}+\bm{\delta})
  +\mathcal{E}_1(\bm{\lambda})\,,
  \label{rat_CS_shapeinv}
\end{equation}
where $\bm{\delta}$ stands for a set of constants indicating the
elementary shift in the parameters.
Isospectral Hamiltonians can be constructed in the following way.\footnote{
Isospectral Hamiltonians can be constructed {\em without} shape invariance
and some formulas e.g., \eqref{rat_CS_phi=Aphi} and \eqref{rat_CS_phi=Adphi},
are also valid in that case. However, there is no explicit systematic method
to obtain $\mathcal{A}_s$ (and $\mathcal{E}_n$) in general.
} 
Starting from $\mathcal{A}_0=\mathcal{A}$, $\mathcal{H}_0=\mathcal{H}$
and $\phi_{0,n}=\phi_n$, let us define $\mathcal{A}_s$, $\mathcal{H}_s$
and $\phi_{s,n}$ ($n\geq s\geq 0$) recursively:
\begin{align}
  \mathcal{A}_{s+1}(y\,;\bm{\lambda})&\eqdef
  \mathcal{A}_s(y\,;\bm{\lambda}+\bm{\delta})\,,
  \label{rat_CS_A_rec}\\
  \mathcal{H}_{s+1}(y\,;\bm{\lambda})&\eqdef
  \mathcal{A}_s(y\,;\bm{\lambda})\mathcal{A}_s(y\,;\bm{\lambda})^{\dagger}
  +\mathcal{E}_s(\bm{\lambda})\,,\\
  \phi_{s+1,n}(y\,;\bm{\lambda})&\eqdef
  \mathcal{A}_s(y\,;\bm{\lambda})\phi_{s,n}(y\,;\bm{\lambda})\,.
  \label{rat_CS_phi_rec}
\end{align}
As a consequence of the shape invariance 
\eqref{rat_CS_shapeinv}, we obtain for $n\geq s\geq 0$
\begin{align}
  &\mathcal{A}_s(y\,;\bm{\lambda})
  =\mathcal{A}(y\,;\bm{\lambda}+s\bm{\delta})\,,
  \label{rat_CS_As}\\
  &\mathcal{H}_s(y\,;\bm{\lambda})
  =\mathcal{A}_s(y\,;\bm{\lambda})^{\dagger}\mathcal{A}_s(y\,;\bm{\lambda})
  +\mathcal{E}_s(\bm{\lambda})
  =\mathcal{H}(y\,;\bm{\lambda}+s\bm{\delta})+\mathcal{E}_s(\bm{\lambda})\,,
  \label{rat_CS_H_rec}\\
  &\mathcal{E}_{s+1}(\bm{\lambda})
  =\mathcal{E}_s(\bm{\lambda})+\mathcal{E}_1(\bm{\lambda}+s\bm{\delta})
  \quad\Bigl(\Rightarrow\mathcal{E}_n(\bm{\lambda})
  =\sum_{s=0}^{n-1}\mathcal{E}_1(\bm{\lambda}+s\bm{\delta})\Bigr)\,,
  \label{rat_CS_E_rec}\\
  &\mathcal{H}_s(y\,;\bm{\lambda})\phi_{s,n}(y\,;\bm{\lambda})
  =\mathcal{E}_n(\bm{\lambda})\phi_{s,n}(y\,;\bm{\lambda})\,,\\
  &\mathcal{A}_s(y\,;\bm{\lambda})\phi_{s,s}(y\,;\bm{\lambda})=0\,,\\
  &\mathcal{A}_s(y\,;\bm{\lambda})^{\dagger}\phi_{s+1,n}(y\,;\bm{\lambda})
  =\bigl(\mathcal{E}_n(\bm{\lambda})-\mathcal{E}_s(\bm{\lambda})\bigr)
  \phi_{s,n}(y\,;\bm{\lambda})\,.
  \label{rat_CS_Adphi}
\end{align}
As is clear from (\ref{rat_CS_H_rec}), all the Hamiltonians 
$\mathcal{H}_0$, $\mathcal{H}_1$, \ldots, $\mathcal{H}_s$, \ldots 
have the same shape and only the parameters are shifted. 
{}From \eqref{rat_CS_phi_rec} and \eqref{rat_CS_Adphi} we obtain
formulas relating the eigenfunctions along the horizontal line (the 
{\em isospectral line}) of Fig.~1,
\begin{align}
  \phi_{s,n}(y\,;\bm{\lambda})&=\mathcal{A}_{s-1}(y\,;\bm{\lambda})\cdots
  \mathcal{A}_1(y\,;\bm{\lambda})\mathcal{A}_0(y\,;\bm{\lambda})
  \phi_n(y\,;\bm{\lambda})\,,
  \label{rat_CS_phi=Aphi}\\
  \phi_n(y\,;\bm{\lambda})&=
  \frac{\mathcal{A}_0(y\,;\bm{\lambda})^{\dagger}}
       {\mathcal{E}_n(\bm{\lambda})-\mathcal{E}_0(\bm{\lambda})}\,
  \frac{\mathcal{A}_1(y\,;\bm{\lambda})^{\dagger}}
       {\mathcal{E}_n(\bm{\lambda})-\mathcal{E}_1(\bm{\lambda})}\cdots
  \frac{\mathcal{A}_{n-1}(y\,;\bm{\lambda})^{\dagger}}
       {\mathcal{E}_n(\bm{\lambda})-\mathcal{E}_{n-1}(\bm{\lambda})}\,
  \phi_{n,n}(y\,;\bm{\lambda})\,,
  \label{rat_CS_phi=Adphi}
\end{align}
and from \eqref{rat_CS_H_rec} we have
\begin{equation}
  \phi_{n,n}(y\,;\bm{\lambda})\propto\phi_0(y\,;\bm{\lambda}+n\bm{\delta}).
  \label{rat_CS_phi=phi}
\end{equation}
It should be emphasized that all the operators $\mathcal{A}$ and
$\mathcal{A}^\dagger$ in the above formulas are explicitly known
thanks to the shape invariance.
The formula \eqref{rat_CS_phi=Adphi} with \eqref{rat_CS_phi=phi}
can be understood as the Rodrigues-type formula.
The relation \eqref{rat_CS_E_rec} implies that
$\{\mathcal{E}_n(\bm{\lambda})\}_{n\geq 0}$ can be calculated from
$\mathcal{E}_1(\bm{\lambda})$, and hence that the spectrum is determined
completely by the shape invariance.

\begin{figure}
  \centerline{\includegraphics[width=8cm]{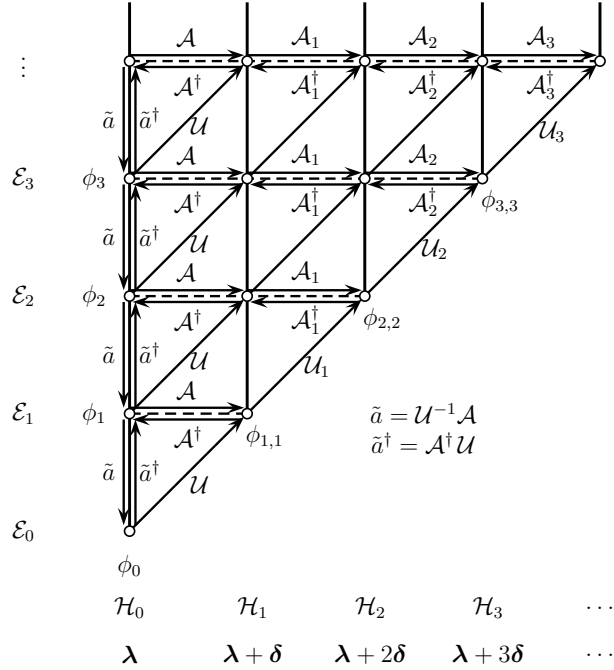}}
  \caption{A schematic diagram of the energy levels and the associated
  Hamiltonian systems  together with the definition of the $\mathcal{A}$
  and $\mathcal{A}^\dagger$ operators and the `creation'
  ($\tilde{a}^{\dagger}$) and `annihilation' ($\tilde{a}$) operators.
  The parameter set is indicated below each Hamiltonian.}
\end{figure}

As seen above, the operators $\mathcal{A}$ and $\mathcal{A}^\dagger$ act
isospectrally, that is, horizontally in Fig.~1. 
On the other hand, the annihilation and creation operators map from one 
eigenstate to another ({\em i.e.\/} vertically) of a given Hamiltonian.
In order to define the annihilation and creation operators, we need 
unitary operators $\mathcal{U}_s$. (For details see for example 
Refs.~\citen{SVZ93,KD02,OS4}.)

The scheme described above is illustrated in Fig.~1.

\subsection{`Discrete' quantum mechanics}
\label{App:DQM}

Here we discuss two different types of shape invariance mechanisms. 
The first, with an additive shift of the parameters, appears 
in the ordinary quantum mechanics discussed in the preceding subsection.
The second one, involving a multiplicative shift of the parameters, is new.

\subsubsection{Additive shift}
First, let us consider the Hamiltonian
\begin{equation}
  \mathcal{H}=\frac12\Bigl(
  \sqrt{V(y)}\,e^{-i\frac{d}{dy}}\sqrt{V(y)^*}
  +\sqrt{V(y)^*}\,e^{i\frac{d}{dy}}\sqrt{V(y)}
  -V(y)-V(y)^*\Bigr)
  \label{rat_RS_calH}
\end{equation}
and its eigenfunctions given in \eqref{cHphi=cEphi}. 
The function $V(y)$ depends on a set of parameters $\bm{\lambda}$.
This Hamiltonian $\mathcal{H}$ is factorizable:
\begin{align}
  \mathcal{H}&=\mathcal{H}(y\,;\bm{\lambda})
  =\mathcal{A}(y\,;\bm{\lambda})^{\dagger}\mathcal{A}(y\,;\bm{\lambda})\,,
  \label{rat_RS_calH=AA}\\
  \mathcal{A}&=\mathcal{A}(y\,;\bm{\lambda})\eqdef\frac{1}{\sqrt{2}}\Bigl(
  e^{-\frac{i}{2}\frac{d}{dy}}\sqrt{V(y\,;\bm{\lambda})^*}
  -e^{\frac{i}{2}\frac{d}{dy}}\sqrt{V(y\,;\bm{\lambda})}\,\Bigr)\,,\\
  \mathcal{A}^{\dagger}&=\mathcal{A}(y\,;\bm{\lambda})^{\dagger}
  \eqdef\frac{1}{\sqrt{2}}\Bigl(
  \sqrt{V(y\,;\bm{\lambda})}\,e^{-\frac{i}{2}\frac{d}{dy}}
  -\sqrt{V(y\,;\bm{\lambda})^*}\,e^{\frac{i}{2}\frac{d}{dy}}\Bigr)\,.
  \label{rat_RS_Ad}
\end{align}
The ground state $\phi_0$ is annihilated by $\mathcal{A}$,
$\mathcal{A}\phi_0=0$.

As in \S\ref{App:QM} we assume the shape invariance
\eqref{rat_CS_shapeinv}; that is,
$\mathcal{A}\mathcal{A}^\dagger$ has the same shape as 
$\mathcal{A}^\dagger\mathcal{A}$ with the additive shift of the
parameters $\bm{\lambda}$ combined with the excitation energy of the
first level $\mathcal{E}_1(\bm{\lambda})$.
Then the isospectral Hamiltonians can be constructed in the same way 
as in the previous subsection and
equations \eqref{rat_CS_A_rec}--\eqref{rat_CS_phi=phi} also hold
here.\cite{OS4}

\subsubsection{Multiplicative shift}
Next let us consider the Hamiltonian
\begin{equation}
  \mathcal{H}=\frac12\Bigl(\sqrt{V(z)}\,q^{D_z}\sqrt{V(z)^*}
  +\sqrt{V(z)^*}\,q^{-D_z}\sqrt{V(z)}-V(z)-V(z)^*\Bigr)
  \label{trig_RS_calH}
\end{equation}
and its eigenfunctions \eqref{cHphi=cEphi}.
The variable $z$ is related to $y$ as $z=e^{2iy}$ and $D_z=z\frac{d}{dz}$.
(In our notation, $V(z)$ here corresponds to $V(y)^*$ in \eqref{rat_RS_calH}.)
This form is more convenient for the trigonometric potentials.
The function $V(z)$ depends on a set of parameters $\bm{\lambda}$ and a
parameter $q$, which controls the multiplicative shift.
This Hamiltonian $\mathcal{H}$ is factorizable:
\begin{align}
  \mathcal{H}&=\mathcal{H}(z\,;\bm{\lambda},q)
  =\mathcal{A}(z\,;\bm{\lambda},q)^{\dagger}\mathcal{A}(z\,;\bm{\lambda},q)\,,
  \label{trig_RS_calH=AA}\\
  \mathcal{A}&=\mathcal{A}(z\,;\bm{\lambda},q)\eqdef\frac{1}{\sqrt{2}}\Bigl(
  q^{\frac12D_z}\sqrt{V(z\,;\bm{\lambda},q)^*}
  -q^{-\frac12D_z}\sqrt{V(z\,;\bm{\lambda},q)}\,\Bigr)\,,\\
  \mathcal{A}^{\dagger}&=\mathcal{A}(z\,;\bm{\lambda},q)^{\dagger}
  \eqdef\frac{1}{\sqrt{2}}\Bigl(
  \sqrt{V(z\,;\bm{\lambda},q)}\,q^{\frac12D_z}
  -\sqrt{V(z\,;\bm{\lambda},q)^*}\,q^{-\frac12D_z}\Bigr)\,.
\end{align}
The ground state $\phi_0$ is annihilated by $\mathcal{A}$,
$\mathcal{A}\phi_0=0$.

We assume the following type of shape invariance
\begin{equation}
  \mathcal{A}(z\,;\bm{\lambda},q)\mathcal{A}(z\,;\bm{\lambda},q)^{\dagger}
  =q^{2\delta'}\mathcal{A}(z\,;q^{\delta}\bm{\lambda},q)^{\dagger}
  \mathcal{A}(z\,;q^{\delta}\bm{\lambda},q)+\mathcal{E}_1(\bm{\lambda},q)\,,
  \label{trig_RS_shapeinv}
\end{equation}
where $\delta$ and $\delta'$ are constants.
As in \S\ref{App:QM} isospectral Hamiltonians can be
constructed.\cite{OS5} \ 
Starting from $\mathcal{A}_0=\mathcal{A}$, $\mathcal{H}_0=\mathcal{H}$
and $\phi_{0,n}=\phi_n$, let us define $\mathcal{A}_s$, $\mathcal{H}_s$
and $\phi_{s,n}$ ($n\geq s\geq 0$) recursively as
\begin{align}
  \mathcal{A}_{s+1}(z\,;\bm{\lambda},q)&\eqdef
  q^{\delta'}\mathcal{A}_s(z\,;q^{\delta}\bm{\lambda},q)\,,
  \label{trig_RS_A_rec}\\
  \mathcal{H}_{s+1}(z\,;\bm{\lambda},q)&\eqdef
  \mathcal{A}_s(z\,;\bm{\lambda},q)\mathcal{A}_s(z\,;\bm{\lambda},q)^{\dagger}
  +\mathcal{E}_s(\bm{\lambda},q)\,,\\
  \phi_{s+1,n}(z\,;\bm{\lambda},q)&\eqdef
  \mathcal{A}_s(z\,;\bm{\lambda},q)\phi_{s,n}(z\,;\bm{\lambda},q)\,.
  \label{trig_RS_phi_rec}
\end{align}
As a consequence of the multiplicative shape invariance
\eqref{trig_RS_shapeinv}, we obtain for $n\geq s\geq 0$
\begin{align}
  &\mathcal{A}_s(z\,;\bm{\lambda},q)=q^{s\delta'}
  \mathcal{A}(z\,;q^{s\delta}\bm{\lambda},q)\,,\\
  &\mathcal{H}_s(z\,;\bm{\lambda},q)
  =\mathcal{A}_s(z\,;\bm{\lambda},q)^{\dagger}\mathcal{A}_s(z\,;\bm{\lambda},q)
  +\mathcal{E}_s(\bm{\lambda},q)\n
  &\phantom{\mathcal{H}_s(z\,;\bm{\lambda},q)}
  =q^{2s\delta'}\mathcal{H}(z\,;q^{s\delta}\bm{\lambda},q)
  +\mathcal{E}_s(\bm{\lambda},q)\,,
  \label{trig_RS_H_rec}\\
  &\mathcal{E}_{s+1}(\bm{\lambda},q)
  =\mathcal{E}_s(\bm{\lambda},q)
  +q^{2s\delta'}\mathcal{E}_1(q^{s\delta}\bm{\lambda},q)
  \ \Bigl(\Rightarrow\mathcal{E}_n(\bm{\lambda},q)
  =\sum_{s=0}^{n-1}q^{2s\delta'}
  \mathcal{E}_1(q^{s\delta}\bm{\lambda},q)\Bigr),
  \label{trig_RS_E_rec}\\
  &\mathcal{H}_s(z\,;\bm{\lambda},q)\phi_{s,n}(z\,;\bm{\lambda},q)
  =\mathcal{E}_n(\bm{\lambda},q)\phi_{s,n}(z\,;\bm{\lambda},q)\,,\\
  &\mathcal{A}_s(z\,;\bm{\lambda},q)\phi_{s,s}(z\,;\bm{\lambda},q)=0\,,\\
  &\mathcal{A}_s(z\,;\bm{\lambda},q)^{\dagger}\phi_{s+1,n}(z\,;\bm{\lambda},q)
  =\bigl(\mathcal{E}_n(\bm{\lambda},q)-\mathcal{E}_s(\bm{\lambda},q)\bigr)
  \phi_{s,n}(z\,;\bm{\lambda},q)\,.
  \label{trig_RS_Adphi}
\end{align}
It should be noted that the Hamiltonian $\mathcal{H}$ is rescaled by a
factor $q^{2\delta'}$ and the parameters $\bm{\lambda}$
are multiplied by a factor $q^\delta$ at each step.
{}From \eqref{trig_RS_phi_rec} and \eqref{trig_RS_Adphi}
we obtain the formulas
\begin{align}
  \phi_{s,n}(z\,;\bm{\lambda},q)&=\mathcal{A}_{s-1}(z\,;\bm{\lambda},q)\cdots
  \mathcal{A}_1(z\,;\bm{\lambda},q)
  \mathcal{A}_0(z\,;\bm{\lambda},q)\phi_n(z\,;\bm{\lambda},q)\,,\\
  \phi_n(z\,;\bm{\lambda},q)&=
  \frac{\mathcal{A}_0(z\,;\bm{\lambda},q)^{\dagger}}
       {\mathcal{E}_n(\bm{\lambda},q)-\mathcal{E}_0(\bm{\lambda},q)}\,
  \frac{\mathcal{A}_1(z\,;\bm{\lambda},q)^{\dagger}}
       {\mathcal{E}_n(\bm{\lambda},q)-\mathcal{E}_1(\bm{\lambda},q)}\cdots\n
  &\qquad\qquad\qquad\times\cdots
  \frac{\mathcal{A}_{n-1}(z\,;\bm{\lambda},q)^{\dagger}}
       {\mathcal{E}_n(\bm{\lambda},q)-\mathcal{E}_{n-1}(\bm{\lambda},q)}\,
  \phi_{n,n}(z\,;\bm{\lambda},q)\,,
  \label{trig_RS_phi=Adphi}
\end{align}
and from \eqref{trig_RS_H_rec} we have
\begin{equation}
  \phi_{n,n}(z\,;\bm{\lambda},q)\propto
  \phi_0(z\,;q^{n\delta}\bm{\lambda},q).
  \label{trig_RS_phi=phi}
\end{equation}


\end{document}